\documentclass[preprintnumbers,eqsecnum,aps,prd,longbibliography,notitlepage,twocolumn]{revtex4-1}
\usepackage{graphicx}
\usepackage{latexsym}
\usepackage{amsmath}
\usepackage{amssymb}
\usepackage{color}
\usepackage[usenames,dvipsnames,svgnames,table]{xcolor}
\usepackage{hyperref}

\def\beq{\begin{equation}}
\def\eeq{\end{equation}}

\newcommand{\nn}{\nonumber}
\newcommand{\mc}[1]{\mathcal{#1}}

\allowdisplaybreaks[1]

\begin{document}

\title{Gravitational self-force corrections to gyroscope precession along circular orbits in the Kerr spacetime}

\author{Donato Bini}
\affiliation{
Istituto per le Applicazioni del Calcolo ``M. Picone,'' CNR, I-00185 Rome,
Italy and \\
National Institute of Nuclear Physics, Section of Roma Tre, Rome, Italy}

\author{Thibault Damour}
  \affiliation{Institut des Hautes \'{E}tudes Scientifiques, F-91440
    Bures-sur-Yvette, France}

\author{Andrea Geralico}
  \affiliation{
Istituto per le Applicazioni del Calcolo ``M. Picone,'' CNR, I-00185 Rome,
Italy}

\author{Chris Kavanagh}
  \affiliation{Institut des Hautes \'{E}tudes Scientifiques, F-91440
    Bures-sur-Yvette, France}

\author{Maarten van de Meent}
\affiliation{Max Planck Institute for Gravitational Physics (Albert Einstein Institute), Potsdam-Golm, Germany and\\
Mathematical Sciences, University of Southampton, United Kingdom}

\date{\today}

\begin{abstract}
We generalize to Kerr spacetime previous gravitational self-force results on gyroscope precession along circular orbits in the Schwarzschild spacetime. In particular we present high order post-Newtonian expansions for the gauge invariant precession function along circular geodesics valid for arbitrary Kerr spin parameter and show agreement between these results and those derived from the full post-Newtonian conservative dynamics. Finally we present strong field numerical data for a range of the Kerr spin parameter, showing agreement with the GSF-PN results, and the expected lightring divergent behaviour. These results provide useful testing benchmarks for self-force calculations in Kerr spacetime, and provide an avenue for translating self-force data into the spin-spin coupling in effective-one-body models.

\end{abstract}

\maketitle

\section{Introduction}

The discovery of gravitational wave signals~\cite{Abbott:2016blz,Abbott:2016nmj,Abbott:2017oio,TheLIGOScientific:2017qsa} associated with the coalescence of two gravitationally interacting compact bodies (either black holes or neutron stars) and the ongoing analysis of the signals has demonstrated the importance of having accurate mathematical descriptions of the underlying dynamics. Hence, updating such models with useful information coming from different (analytic, semi-analytic or numeric) approximation methods remains an active research area. Spurring this on further is the promise of a wide range of complicated low-frequency gravitational wave sources visible by the space based interferometer LISA \cite{Klein:2015hvg,Caprini:2015zlo,Tamanini:2016zlh,Bartolo:2016ami,Babak:2017tow}

All existing methods have indeed a limited range of applicability. For example, when the two-body dynamics occurs in a weak-field and slow motion regime the key method is the post-Newtonian (PN) expansion; when, instead, the field is weak but the motion is no longer slow one can apply the post-Minkowskian (PM) approximation; finally, when the mass-ratio of the two bodies is very small the general relativistic perturbation theory on the field of the large mass---referred to as the gravitational self-force approach (GSF in short)---can be conveniently used. In addition to these more analytic approaches, there is numerical relativity (NR) where one directly solves the full Einstein equations numerically without making any fundamental approximations. While NR offers the only direct approach to viewing the merger and ringdown phases of a binary coalescence, computational costs exclude the early inspiral and situations where the binary has a small mass ratio.
These methods have been developed independently from each other, allowing fruitful crosschecking of results.

This paper is concerned with phenomena associated with extreme mass-ratio inspirals, a source for LISA which is most naturally described using GSF. Of particular interest has been the identification of gauge invariant physical effects of the conservative self-force, for example the well known periastron advance \cite{Barack:2010ny,
vandeMeent:2016hel} and redshift invariants \cite{Detweiler:2008ft,Barack:2011ed,Sago:2008id,Shah:2010bi,Shah:2012gu,vandeMeent:2015lxa,Kavanagh:2016idg,Kavanagh:2015lva,Bini:2016dvs}; for a recent review of this topic see \cite{Barack:2018yvs}. These invariants rely on the delicate regularization techniques for dealing with the singular nature of the point like source. Therefore, calculating and comparing such invariants with either other independent  self-force codes, post-Newtonian methods or numerical relativity simulations, one can confirm difficult calculations and validate new codes. The development of conservative gauge invariants has probed ever higher derivatives of the metric perturbations, requiring more careful regularization. We summarise the current knowledge of these invariants in Table~\ref{tab:GaugeInvariants}.

The aim of the present work is to expand the range of gauge invariants to include knowledge of the GSF corrections to the accumulated precession angle of the spin vector of a test gyroscope per radian of orbital motion, commonly referred to as the spin precession invariant. The gyroscope (carrying a small mass $m_1$ and a small spin $S_1$) moves along a circular geodesic orbit in Kerr spacetime (with mass $m_2$, spin $S_2$), generalizing previous results for a nonrotating black hole \cite{Dolan:2013roa,Bini:2014ica}. We see this also as a basis for generalizing the more difficult eccentric orbit calculation in Schwarzschild spacetime \cite{Akcay:2016dku,Kavanagh:2017wot} to Kerr spacetime as suggested in \cite{Akcay:2017azq}. 
We will use the notation $a_1=S_1/m_1$ and $a_2=S_2/m_2$ for the spin-to-mass ratio of the two bodies and associated dimensionless spin variables $\chi_1=S_1/m_1^2$ and $\chi_2=S_2/m_2^2$.
Other standard notations are $M=m_1+m_2$ for the total mass of the system
and
\beq
q=\frac{m_1}{m_2}\ll 1 \,,\qquad \nu=\frac{m_1 m_2}{(m_1+m_2)^2}\ll1 \,,
\eeq
for the ordinary and symmetric mass-ratios, respectively. See Table~\ref{tab:notation} for an overview of our notational conventions.

Unless differently specified we will use units so that $c=G=1$.

\begin{table}[htt]
	\caption{Overview of calculations of the various gauge invariants in the literature and a sample of references. }
	\begin{ruledtabular}
		\begin{tabular}{lcc}
			 & \textbf{Schwarzschild}  & \textbf{Kerr}  \\
			\hline
			Redshift  	& \checkmark \cite{Detweiler:2008ft,Sago:2008id,Shah:2010bi,Barack:2011ed,Bini:2013zaa,Bini:2013rfa, Bini:2014nfa,Kavanagh:2015lva,Hopper:2015icj,Johnson-McDaniel:2015vva,Bini:2015bla,Bini:2015bfb,Bini:2016qtx,Bini:2018zde} &   \checkmark	\cite{Shah:2012gu,Shah:2013uya,vandeMeent:2015lxa,Bini:2015xua,Kavanagh:2016idg,Bini:2016dvs}	\\
			Spin precession  	&  \checkmark \cite{Dolan:2013roa,Bini:2014ica,Bini:2015mza,Kavanagh:2015lva,Shah:2015nva,Akcay:2016dku,Kavanagh:2017wot,Bini:2018aps} &  	This work	\\
			Quadrupolar Tidal  	&  \checkmark \cite{Dolan:2014pja,Bini:2014zxa,Bini:2014ica,Kavanagh:2015lva,Shah:2015nva,Bini:2018kov} &   \checkmark 	\cite{Bini:2018dki}\\
			Octupolar Tidal  	&  \checkmark \cite{Bini:2014zxa,Nolan:2015vpa} &   
		\end{tabular}	
	\end{ruledtabular}
\label{tab:GaugeInvariants}
\end{table}

\begin{table}[htt]
	\caption{List of notations related to mass and spin used in this paper. Care is required since notation often varies between GSF and PN literature.}
	\begin{ruledtabular}
		\begin{tabular}{ll}
			 $m_1$ & mass of small body \\
			 $m_2$ & mass of Kerr BH \\
			 $M$ & $m_1+m_2$ \\
			 $q$ & small mass ratio \\
			 $\nu$ & symmetric mass ratio \\
			 $S_i$ & spin magnitude of body $i$ \\
			 $a_i$ & $S_i/m_i$ \\
			 $a$ & $a_2$ \\
			 $\hat{a}$ & $a/m_2$ \\
			 $\chi_i$ & $a_i/m_i$
		\end{tabular}	
	\end{ruledtabular}
\label{tab:notation}
\end{table}

\section{Kerr metric and perturbation}

The (unperturbed) Kerr line element written in standard Boyer-Lindquist coordinates $(t,r,\theta,\phi)$ reads
\begin{eqnarray}
\label{kerrmet}
ds_{(0)}^{2}&=& g^{(0)}_{\alpha\beta} dx^\alpha dx^\beta \nonumber\\
&=&-\left(1-\frac{2m_2r}{\Sigma}  \right) dt^2-\frac{4am_2r \sin^2\theta}{\Sigma}dtd\phi\nonumber\\
&+&\frac{\Sigma}{\Delta}dr^2
+ \Sigma d\theta^2 \nonumber\\
&+&\left( r^2+a^2+\frac{2m_2ra^2\sin^2\theta}{\Sigma} \right)\sin^2\theta d\phi^2\,,
\end{eqnarray}
where   
\beq
\Delta= r^2+a^2-2m_2r\,,\qquad 
\Sigma=r^2+a^2\cos^2\theta\,.
\eeq

Let us consider the perturbation induced by a test gyroscope moving along a circular equatorial orbit at~$r=r_0$.
The perturbed regularized metric will be denoted by $g^{\rm R}_{\alpha\beta}= g^{(0)}_{\alpha\beta}+q h^{\rm R}_{\alpha\beta}+\mathcal{O}(q^2)$, with corresponding line element
\beq
\label{kerrmet_pert}
ds^2=(g^{(0)}_{\alpha\beta}+q h_{\alpha\beta}+\mc{O}(q^2)) dx^\alpha dx^\beta \,,
\eeq
and is assumed to keep a helical symmetry, with associated Killing vector $k=\partial_t +\Omega \partial_\phi$.
Because of the helical symmetry, the metric perturbation depend only on $\bar \phi=\phi-\Omega t$, $r$ and $\theta$, i.e., $h_{\mu\nu}=h_{\mu\nu}(\bar \phi , r, \theta)$. 

The gyroscope world line (in both the unperturbed and perturbed cases) has its unit timelike tangent vector aligned with $k^\alpha$, i.e.,
\beq
u^\alpha= u^tk^\alpha\,,
\eeq
where $u^t$ is a normalization factor (such that $u^\alpha u_\alpha=-1$). 
In the (unperturbed) Kerr case the orbital frequency is given by
\beq
\label{3.4_0}
m_2\Omega^{(0)} = \frac{u^{3/2}}{1+{\hat a}u^{3/2}}\,,
\eeq
and
\beq
u^{t\,(0)}= \frac{1+\hat a u^{3/2}}{\sqrt{1-3u+2\hat a u^{3/2}}}\,,
\eeq
with $u=m_2/r$ the dimensionless inverse radius of the orbit.
In the perturbed situation the frequency becomes 
\begin{align}
\label{3.4}
m_2\Omega &=&m_2\Omega^{(0)}\left(1-q\frac{1+{\hat a}u^{3/2}}{4u^2} m_2[\partial_r h_{kk}^{\rm R}]_1+\mc{O}(q^2)\right) \,,
\end{align}
where we have denoted as
\beq
\label{3.5}
h_{kk}^{\rm R}=h_{\alpha\beta}^{\rm R}k^\alpha k^\beta\,,
\eeq
the double contraction of $h^{\rm R}$ with the Killing vector $k$, and the subscript 1 stands for the evaluation at the position of the particle 1.
In the perturbed metric the geodesic condition also implies $[\partial_{\bar \phi} h_{kk}^{\rm R}]_1=0$ \cite{Detweiler:2008ft}.

\subsection{Gyroscope precession}

We are interested in computing the spin precession invariant $\psi(y)$, measuring the accumulated precession angle of the spin vector of a test gyroscope per radian of orbital motion defined in \cite{Dolan:2013roa}. That is, the ratio of the \lq\lq geodetic" spin precession frequency to the orbital frequency
\beq
\label{5.1}
\psi(y)\equiv \frac{\Omega_{\rm prec}}{\Omega},
\eeq 
where $\psi$ is written as a function of the gauge-invariant dimensionless frequency parameter
\beq
\label{3.6}
y=(m_2\Omega)^{2/3}\,.
\eeq
We thus compute the precession frequency $\Omega_{\rm prec}$ of the small-mass body 1 carrying a small-spin orbiting the large-mass spinning body 2, to linear order in the mass-ratio.
The precession frequency, both in the background and in the perturbed spacetime, is defined by (see, e.g., Ref. \cite{Bini:2014ica})
\beq
\label{1.12}
\Omega_{\rm prec}=\Omega-|\nabla k|\,,
\eeq
where 
\beq
|\nabla k|^2=\frac12 [K_{\mu\nu}K^{\mu\nu}]_1\,, 
\eeq
with 
\beq
\label{1.4}
K_{\mu\nu}=\nabla_\mu^{\rm R} k_\nu=-\nabla_\nu^{\rm R} k_\mu=\frac12 (\partial_\mu k_\nu-\partial_\nu k_\mu)\,.
\eeq

In terms of the gauge-invariant dimensionless frequency parameter \eqref{3.6},
Eq. \eqref{3.4} implies
\beq
u = \frac{y}{(1-{\hat a}y^{3/2})^{2/3}}\left(1+q\frac{m_2[\partial_r h_{kk}^{\rm R}]_1}{6y^2(1-{\hat a}y^{3/2})^{2/3}}+\mc{O}(q^2)\right)\,.
\eeq
We then have
\beq
\label{3.9}
m_2 |\nabla k|= m_2 |\nabla k|^{(0)}\, (1+q\, \delta(y)+\mathcal{O}(q^2))\,,
\eeq
where
\beq
m_2 |\nabla k|^{(0)}=y^{3/2}\left[\frac{1+{\hat a}y^{3/2}-3y\left(1-{\hat a}y^{3/2}\right)^{1/3}}{1-{\hat a}y^{3/2}}\right]^{1/2}\,,
\eeq
and
\begin{eqnarray}
\delta(y)&=&
\frac{\hat a}{2u^{1/2}}(1+{\hat a}u^{3/2})m_2\partial_r h_{kk}^{\rm R}\nonumber\\
&&+\frac1{2u^{1/2}} \left(\partial_r h_{\phi k}^{\rm R}-\partial_{\phi} h_{r k}^{\rm R}\right)\nonumber\\
&&
-\frac{u}2 \frac{(1-2{\hat a}u^{3/2}+{\hat a}^2u^2)^2}{(1-3u+2{\hat a}u^{3/2})(1-2u+{\hat a}^2u^2)} h_{kk}^{\rm R}\nn\\
&&
-u^{3/2}\frac{1-2{\hat a}u^{3/2}+{\hat a}^2u^2}{(1-2u+{\hat a}^2u^2)(1+{\hat a}u^{3/2})}\frac{1}{m_2}h_{t\phi}^{\rm R}
\nonumber\\
&&
-\frac{u^2}2  \frac{1-u+2{\hat a}u^{3/2}(1-2u)+2{\hat a}^2u^3}{(1-2u+{\hat a}^2u^2)(1+{\hat a}u^{3/2})^2}\frac{1}{m_2^2}h_{\phi\phi}^{\rm R}\nn\\
&&
-\frac12 (1-2u+{\hat a}^2u^2)h_{rr}^{\rm R}\,,\label{eq:delta}
\end{eqnarray}
with $u = {y}/{(1-{\hat a}y^{3/2})^{2/3}}$ in this (and in any) $\mathcal{O}(q)$ quantity.

To linear order in $q$ (keeping $y$ as fixed), inserting \eqref{3.9} into \eqref{5.1}, $\psi(y)$ reads
\beq
\label{5.2}
\psi (y)=1-\frac{|\nabla k|^{(0)}}{y^{3/2}}[1+q\, \delta(y)+\mathcal{O}(q^2)]\,,
\eeq 
where
\begin{align}
	\frac{|\nabla k|^{(0)}}{y^{3/2}}=\left[\frac{1+{\hat a}y^{3/2}-3y\left(1-{\hat a}y^{3/2}\right)^{1/3}}{1-{\hat a}y^{3/2}}\right]^{1/2}\,,
\end{align}
so that the GSF piece $\delta \psi (y)$ in $\psi (y)$ (such that $\psi(y)=\psi_{(0)}(y)+q\, \delta \psi (y)+\mathcal{O}(q^2)$) is related to 
$\delta(y)$ via
\beq
\label{5.3}
\delta \psi (y)=-\frac{|\nabla k|^{(0)}}{y^{3/2}}\, \delta(y)\,.
\eeq 
The well known (unperturbed) Kerr result is then recovered, namely
\begin{eqnarray}
m_2\Omega_{\rm prec}^{(0)}&=& m_2\Omega^{(0)}-m_2|\nabla k|^{(0)}\\
&=& y^{3/2}\left[1-\sqrt{\frac{1+\hat a y^{3/2}}{1-\hat a y^{3/2}}-3y(1-\hat a y^{3/2})^{-2/3}}  \right]\,,\nonumber
\end{eqnarray}
together with the corresponding Schwarzschild limit ($\hat a\rightarrow0$)
\begin{eqnarray}
m_2\Omega_{\rm prec,\, schw}^{(0)}&=& 
y^{3/2}\left[1-\sqrt{1-3y}  \right]\,.
\end{eqnarray}
Our goal for the remainder of the paper is to evaluate \eqref{5.3} using \eqref{eq:delta}.

\section{Methods}

\begin{table}
	\caption{Overview of the different implementations used in this work.
The two analytic methods I and II differ for the field mode decomposition in terms of either scalar spherical harmonics $Y_{lm}$ or spin-weighted spheroidal harmonics ${}_s S_{lm}$ and their respective derivatives.
The third approach III, instead, uses purely a scalar harmonic projection. 
}
	\begin{ruledtabular}
		\begin{tabular}{cccccc}
			\# & Weyl scalar & Gauge & Mode decomposition& Method & Ref \\
			\hline
			I  	& $\psi_0$ &  ORG&	${}_s S_{lm},\partial_\theta{}_s S_{lm}	$& analytic	 & \cite{Kavanagh:2016idg}		\\
			II  	& $\psi_0$ &  ORG&	  $Y_{lm},\partial_\theta Y_{lm}$  & analytic	& \cite{Bini:2016dvs}			\\
			III	& $\psi_4$ & ORG&	$Y_{lm}	$& numeric & \cite{vandeMeent:2015lxa}
		\end{tabular}
	\end{ruledtabular}
\label{tab:1}
\end{table}

\subsection{Radiation gauge metric reconstruction}

In this work we will follow the Chrzanowski-Cohen-Kegeles (CCK) procedure for obtaining metric perturbations in a radiation gauge \cite{Cohen:1974cm,Chrzanowski:1975wv,Kegeles:1979an,Wald:1978vm,Lousto:2002em,Ori:2002uv,Keidl:2010pm,Shah:2012gu,vandeMeent:2015lxa}.
Once a solution for the perturbed Weyl scalar $\psi_0$ (or $\psi_4$) has been obtained by solving the $s=2$ (or $s=-2$) Teukolsky equation, one then construct the Hertz potential $\hat{\Psi}_{0/4}$, in terms of which one finally compute the components of the perturbed metric $h_{\alpha\beta}^{\mathrm{rec}}$ by applying a suitable differential operator.
We will use the outgoing radiation gauge (ORG) (see Table \ref{tab:1}), such that the metric perturbation $h_{\alpha\beta}^{\mathrm{rec}}$ satisfies the conditions
\beq
n^\alpha h^{\mathrm{rec}}_{\alpha\beta} = 0,\qquad
{h^{\mathrm{rec}}}_{\alpha}^{\phantom{\alpha}\alpha} =0,
\eeq
where $n^\alpha$ is the ingoing principal null vector.

When sources are present, radiation gauge solutions to the Einstein equations feature singularities away from the source region \cite{Ori:2002uv,Barack:2001ph}. If the source is a point particle, a string-like (gauge) singularity will extend from the particle to infinity and/or the background horizon \cite{Pound:2013faa,Keidl:2006wk}. Alternatively, one can construct a solution obtained by gluing together the regular halves from two `half-string' solutions. The result is a metric perturbation with a gauge discontinuity on a hypersurface containing the point source's worldline.  We work with this solution. The gauge discontinuity splits the spacetime in two disjoint regions: an `exterior' region that extends to infinity (labelled ``$+$''), and an `interior' region that includes the background horizon (labelled ``$-$''). 

All the necessary steps to perform this kind of computations are now well established in the literature (see, e.g., Refs. \cite{Keidl:2010pm,Shah:2012gu}). In this work we implement the CCK procedure using three separate codes, two using analytic methods resulting in high order post-Newtonian expansions of the metric perturbation and thus precession invariant, and one numerical code giving high accuracy data over a finite set of radii and Kerr spin values. We highlight the variations of these methods we follow and provide references for more details of their techniques in Table~\ref{tab:1}.

An aspect in which our methods differ is the basis of angular harmonics in which our fields are represented. Method I keeps the natural basis of spin-weighted spheroidal harmonics for the representation of the Weyl scalars, resulting in an expression for the metric perturbation and spin precession invariant in a combination of  spin-weighted spheroidal harmonics and their angular derivatives. In method II the spin-weighted spheroidal harmonics are expanded in scalar spherical harmonics, resulting in expressions which are a combination of scalar spherical harmonics and their derivatives. In both methods I and II, as a result of the CCK procedure the coefficients in the harmonic expansion can still depend on the angular variables. In method III \emph{all} angular dependence is projected onto an expansion in scalar spherical harmonics.

In all methods we use the solutions of the radial Teukolsky equation due to Mano, Suzuki and Takasugi~(MST) \cite{Mano:1996gn,Mano:1996vt} satisfying the correct boundary conditions at the horizon and at spatial infinity. In methods I and II these are expanded as an asymptotic series in $u$ for certain low values of the harmonic $l$ value, and supplement the MST series with a PN type ansatz for all higher values of $l$, obtaining the spin-precesion invariant as a PN expression. Method III instead evaluates the MST solutions numerically following \cite{Fujita:2004rb,Fujita:2009us,Throwethesis}.

\subsection{Regularization}

The quantity $\delta\psi$ is defined in terms of the Detweiler-Whiting regular field \cite{Detweiler:2002mi}, $h_{\mu\nu}^{\rm R}$. In practice this is obtained as the difference,
\begin{equation}
h_{\mu\nu}^{\rm R} = h_{\mu\nu}^{\rm ret} -h_{\mu\nu}^{\rm S},
\end{equation}
between the retarded field $h_{\mu\nu}^{\rm ret}$ and the Detweiler-Whiting singular field  $h_{\mu\nu}^{\rm S}$. Since both $h_{\mu\nu}^{\rm ret}$ and $h_{\mu\nu}^{\rm S}$ are singular on the particle world line, this subtraction cannot be performed there, but requires the introduction of a suitable regulator. We here use a variant of the so called $l$-mode regularization of \cite{Barack:2001gx}. This calls for extending Eqs. \eqref{eq:delta} and \eqref{5.3} to field equations by choosing an extension of the four-velocity $u$ to a field. Eqs. \eqref{eq:delta} and \eqref{5.3} can then be applied separately to  $h_{\mu\nu}^{\rm ret}$ and $h_{\mu\nu}^{\rm S}$, obtaining the fields $\delta\psi^{\rm ret}$ and $\delta\psi^{\rm S}$. The spherical harmonic modes of these fields are then finite, and the necessary subtraction can be done a the level of these modes
\begin{equation}\label{eq:modesum}
\delta\psi^{\rm R} = \sum_{l=0}^{\infty} (\psi^{\rm ret}_l -\psi^{\rm S}_l).
\end{equation}
Conventional $l$-mode regularization procedures continue to calculate $\psi^{\rm S}$ locally near the worldline with chosen gauge and extension of the 4-velocity, yielding an expression for the large $l$ behaviour of  $\psi^{\rm S}_l$,
\begin{equation}
\label{psi_sing}
\psi^{\rm S}_l = \pm L A_\psi + B_\psi + L^{-1} C_\psi + \mathcal{O}(L^{-2}), 
\end{equation}
with $L=2l+1$, and $\pm$ sign depended on the direction from which the worldline is approached. Consequently, the mode-sum \eqref{eq:modesum} can be evaluate as
\begin{equation}\label{eq:modesum2}
\delta\psi^{\rm R} = \sum_{l=0}^{\infty} (\psi^{\rm ret}_l \mp L A_\psi- B_\psi -L^{-1} C_\psi) -D_\psi,
\end{equation}
with
\begin{equation}\label{eq:Dterm}
D_\psi = \sum_{l=0}^{\infty} (\psi^{\rm S}_l \mp L A_\psi- B_\psi -L^{-1} C_\psi).
\end{equation}
We follow a slightly different approach first applied in Refs. \cite{Shah:2010bi} and \cite{Bini:2013zaa}. 

Since $h_{\mu\nu}^{\rm R}$ is a smooth vacuum perturbation the large $l$ behaviour of $\delta\psi^{\rm R}_l$ is expected the be $\mathcal{O}(\exp(-c/L))$, and consequently we can read off the coefficients $A_\psi$, $B_\psi$, and $C_\psi$ from the large $l$ behaviour of $\psi^{\rm ret}_l$, which we can determine either numerically of in a PN expansion. 

However, it is fundamentally impossible to determine the ``D-term'' \eqref{eq:Dterm} from the retarded field alone. In general it will depend on the chosen gauge, extension, and type of harmonic expansion (e.g. scalar, spin-weighted, mixed, ...). For the GSF it is known to vanish for a large class of (regular) gauges and extensions \cite{Heffernan:2012vj}. However, it is known to take non-zero values in the radiation gauge used in this work. In particular, the D-term will be different in the interior and exterior solutions. In \cite{Pound:2013faa} it was shown that for the GSF, the corrections to the D-term relative to the Lorenz gauge cancel when one takes the average of the interior and exterior solutions. This argument extends a much wider class of quantities (at least for suitably chosen extensions) including $\delta\psi$ \cite{gaugecompletion}. Consequently, if $D_\psi$ vanishes in the Lorenz gauge, we can calculate $\delta\psi^{\rm R}$ through
\beq
\delta\psi^{\rm R}=\sum_{l=0}^\infty \left[\frac12 (\delta\psi_l^+ + \delta\psi_l^-)-B_\psi -C_\psi/L\right].
\eeq
In this work, we \emph{conjecture} that $D_\psi$ vanishes in the Lorenz gauge for the chosen extensions and harmonic decompositions. In part, this conjecture will be motivated post-facto by the agreement of our results with standard PN results up to 4PN order.

In methods I and II we find that the expressions for $B_\psi$ and $C_\psi$ agree (despite differences in harmonic decomposition). In particular we find that $C_\psi=0$ and $B_\psi$ is given by 
\begin{align}
B_\psi=\frac{|\nabla k|^{(0)}}{y^{3/2}}B\,,
\end{align}
with
\beq
B=\sum_{n=0}^N {\hat a}^n B^{a^n}(y)\,,
\eeq
and
\begin{eqnarray}
B^{a^0}(y)&=&
\frac{1}{2}y-\frac{1}{4}y^2-\frac{63}{128}y^3-\frac{995}{1024}y^4-\frac{63223}{32768}y^5\nn\\
&&-\frac{126849}{32768}y^6-\frac{16567767}{2097152}y^7-\frac{555080733}{33554432}y^8\nonumber\\
&&-\frac{77104836855}{2147483648}y^9
\,,\nonumber\\
B^{a^1}(y)&=&
-\frac{1}{2}y^{3/2}+\frac{5}{6}y^{5/2}+\frac{277}{384}y^{7/2}+\frac{1385}{1024}y^{9/2}\nn\\
&&+\frac{272245}{98304}y^{11/2}+\frac{1253839}{196608}y^{13/2}+\frac{34614543}{2097152}y^{15/2}\nonumber\\
&&
+\frac{1563825339}{33554432}y^{17/2}+\frac{297179922135}{2147483648}y^{19/2}
\,,\nonumber\\
B^{a^2}(y)&=&
-\frac{1}{4}y^3+\frac{43}{144}y^4-\frac{335}{9216}y^5-\frac{5953}{4096}y^6\nn\\
&&-\frac{2547251}{294912}y^7-\frac{343117}{9216}y^8-\frac{4673348817}{33554432}y^9
\,,\nonumber\\
B^{a^3}(y)&=&
-\frac{1}{16}y^{9/2}+\frac{103757}{82944}y^{11/2}+\frac{2002033}{331776}y^{13/2}\nn\\
&&+\frac{1566715}{65536}y^{15/2}+\frac{891749345}{10616832}y^{17/2}\nonumber\\
&&+\frac{759009677191}{2717908992}y^{19/2}
\,,\nonumber\\
B^{a^4}(y)&=&
-\frac{39}{128}y^5-\frac{1783}{1024}y^6-\frac{158107}{31104}y^7\nn\\
&&-\frac{13809563}{995328}y^8-\frac{76343163}{2097152}y^9
\,\nonumber\\
B^{a^5}(y)&=&\frac{15}{128} y^{11/2}-\frac{299}{1024} y^{13/2}-\frac{1903 }{512}y^{15/2}\nn\\
&&-\frac{103026043 }{5971968}y^{17/2}
\,\nonumber\\
B^{a^6}(y)&=&\frac{765 }{1024}y^7+\frac{39125 }{12288}y^8+\frac{34971 }{4096}y^9
\,\nonumber\\
B^{a^7}(y)&=&-\frac{175}{1024} y^{15/2}+\frac{14395 }{12288}y^{17/2}
\,\nonumber\\
B^{a^8}(y)&=&-\frac{38535 }{32768}y^9\,.
\end{eqnarray}
and higher powers of $\hat{a}$ appear at higher PN orders.

\subsection{Completion}

Since the operator $\hat{\Psi}_{0/4}$ is not injective, its inverse is fundamentally ambiguous up to an element of the kernel of $\hat{\Psi}_{0/4}$. Wald showed that the only (global) vacuum solutions of the linearized Einstein equation in this kernel are perturbations to the mass and angular momentum of the background Kerr spacetime and pure gauge solutions. Hence the full metric perturbation can be written
\begin{equation}
\label{eq:comph}
h_{\alpha\beta}^\pm = h_{\alpha\beta}^{\mathrm{rec},\pm} + (\partial_{m_2} g^{(0)}_{\alpha\beta})\delta M^\pm+(\partial_{S_2} g^{(0)}_{\alpha\beta})\delta J^\pm
 + \nabla_{(\alpha}\xi_{\beta)}^\pm,
\end{equation}
with $\delta M^\pm$ and $\delta J^\pm$ numbers and $\xi_{\beta}^\pm$ gauge vector fields.

It was shown in \cite{Merlin:2016boc,vandeMeent:2017fqk} that for a particle source on a bound geodesic the amplitudes of the mass and angular momentum perturbations are given by
\begin{align}
 \delta M^{-} &=0\,, & \delta J^{-} &=0\,,	\nn\\ 
 \delta M^{+} &= m_2 E\,, & \delta J^{+} &= m_2 L\,,
\end{align}
where $\hat{E}$ and $\hat{L}$ are the specific energy and angular momentum of the particle,
\begin{eqnarray}
\hat{E} &=& \frac{1-2u+\hat a u^{3/2}}{\sqrt{1-3u+2\hat a u^{3/2}}}\,,\nonumber\\
\frac{\hat{L}}{m_2} &=& \frac{1-2\hat a u^{3/2} +\hat a^2 u^2}{\sqrt{u} \sqrt{1-3u+2\hat a u^{3/2}}}\,.
\end{eqnarray}

If $\delta\psi$ were a proper gauge invariant quantity, then we could simply ignore the gauge vectors $\xi^{\alpha}_\pm$. However, $\delta\psi$ (like the orbital frequency) is only a quasi-invariant in the sense of \cite{gaugecompletion}, meaning that it is only invariant under gauge vectors that are bounded in time. We thus have to (partially) fix the gauge contribution to the metric as well. We start this process by noting that the other contributions to the metric perturbation in \eqref{eq:comph} are all bounded in time. Consequently, by restricting our attention to gauges in which $h_{\alpha\beta}$ is bounded in time, we only have to consider gauge vectors that produce bounded metric perturbations. The most general such gauge vector \cite{gaugecompletion,Shah:2015nva} is
\begin{equation}
\xi_\pm=m_2 t\left(A^t_\pm\partial_t+A^\phi_\pm\partial_\phi\right) + o(t).
\end{equation}
Consequently, to uniquely fix the value of $\delta\psi$, we only need to fix the values of $A^t_\pm$ and $A^\phi_\pm$. The $A_{+}^{t/\phi}$ can be fixed by requiring the full metric perturbation $h_{\alpha\beta}^{+}$ to be asymptotically Minkowski, yielding $A_{+}^{t/\phi}=0$. The interior values $A_{-}^{t/\phi}$ can further be fixed by requiring the continuity of suitably chosen quasi-invariant fields constructed from the metric perturbation \cite{gaugecompletion}. For circular equatorial orbits in Kerr this procedure yields,
\begin{align}
\label{VdMleq01}
A_{-}^{t}&=
\frac{[(u^{1/2}\hat a  -2)\hat a u^{3/2}-1] u}{ (1-3 u+2\hat a u^{3/2})^{1/2}(\hat a u^{3/2}+1)}
\,,\\
A_{-}^{\phi}&=
\frac{(u^{1/2}\hat a-2) u^{5/2}}{(1-3 u+2\hat a u^{3/2})^{1/2}(\hat a u^{3/2}+1)} 
\,.
\end{align}
With this the final expression for the GSF contribution to the spin precession invariant becomes
\begin{equation}
\delta\psi = \delta\psi^\mathrm{rec} 
+ \frac12\left(\hat{E} \delta\psi^M + \frac{\hat{L}}{m_2} \delta\psi^J + A^t_{-} \delta\psi^t +  A^\phi_{-} \delta\psi^\phi\right)\,,
\end{equation}
with 
\begin{align}
\delta\psi^M &=\frac{u(1+\hat{a}u^{1/2} - \hat{a}u^{3/2} +\hat{a}^2 u^2 )}{\sqrt{1-3u+2\hat{a}u^{3/2}}}, \\
\delta\psi^J &=-\frac{u^{3/2}(1 - u + \hat{a}u^{3/2} )}{\sqrt{1-3u+2\hat{a}u^{3/2}}}, \\
\delta\psi^t &=-\frac{u^{3/2}(1-\hat{a}u^{1/2})(1 + \hat{a}u^{3/2})}{\sqrt{1-3u+2\hat{a}u^{3/2}}}, \\
\delta\psi^\phi &=-\frac{1+\hat{a} u^{3/2}}{u^{3/2}\sqrt{1-3u+2\hat{a}u^{3/2}}}\nn\\
&\qquad\times(1 - 4 u + 3\hat{a}u^{3/2}-\hat{a}u^{5/2}+\hat{a}^2u^{3} ).
\end{align}

\section{Results}

\subsection{Spin-Exact results to 8PN}

\noindent Omitting the intermediate results of the radiative and completion parts, the output of method I of Table~\ref{tab:1} is the spin-precession invariant written as a PN series in $y$ and $\log y$ with no restriction on the spin of the black hole $\hat{a}$. The results take the form
\begin{align}
  \delta \psi &= 
      c_{1.5}y^{1.5}+ c_2 y^2 + c_{2.5} y^{2.5} + c_3 y^3 + c_{3.5} y^{3.5} + c_{4} y^{4} \nn\\
     &\quad + c_{4.5} y^{4.5} + (c_{5} + c^{\rm ln}_{5} \log y) y^{5} + c_{5.5} y^{5.5}\nn\\
     &\quad+ (c_{6} + c^{\rm ln}_{6} \log y) y^{6}  + (c_{6.5} + c_{6.5}^{\rm ln} \log y) y^{6.5}\nn\\  
     &\quad + (c_{7} + c^{\rm ln}_{7} \log y) y^{7} + (c_{7.5} + c_{7.5}^{\rm ln} \log y)  y^{7.5}\nn\\
     &\quad + (c_{8} + c^{\rm ln}_{8} \log y + c^{\rm ln^2}_{8} \log^2 y) y^{8}  +\mc{O}(y^{8.5}) . \label{Eq:deltapsi8pn}
\end{align}
The coefficients in this expansion are given by
\begin{eqnarray}
c_{1.5} &=& \hat{a}, \quad
c_{2} = 1, \quad
c_{2.5} =0, \quad
c_{3} = -3, \nonumber\\
c_{3.5} &=& \tfrac{16}{3} \hat{a}, \quad
c_{4} =- \tfrac{15}{2}  -  3 \hat{a}^2, \nonumber \\ 
c_{4.5} &=&  \big[ \tfrac{233}{6} - \tfrac{41}{32} \pi^2 \big]\hat{a}+ \hat{a}^3, \nonumber \\ 
c_{5} &=& - \tfrac{6277}{30} -  16 \gamma + \tfrac{20471}{1024} \pi^2 -  \tfrac{496}{15} \log(2) -  \tfrac{163}{9} \hat{a}^2, \nonumber \\ 
c_{5}^{\rm ln} &=& -8,
\end{eqnarray}
\begin{widetext}

\begin{align}
c_{5.5} &=   \big[-\tfrac{89}{15} + \tfrac{248}{5} \gamma +  \tfrac{1585}{1024} \pi^2 + \tfrac{504}{5} \log(2)\big]\hat{a} + 4 \hat{a}^3, \nonumber \\
c_{5.5}^{\rm ln} &= \tfrac{124}{5}\hat{a}, \nonumber \\ 
c_{6} &= -\tfrac{87055}{28}-\tfrac{52}{5} \gamma +\tfrac{653629}{2048} \pi
   ^2+\tfrac{3772 }{105}\log
   (2)-\tfrac{729 }{14}\log (3)-\big[\tfrac{667}{18}+\tfrac{11023}{3072} \pi ^2\big] \hat{a}^2, \nonumber \\ 
c_{6}^{\rm ln} &=-\tfrac{26}{5}, \nonumber \\ 
c_{6.5} &= -\tfrac{26536 }{1575} \pi+
   \big[-\tfrac{6767891}{700}-\tfrac{3736 }{35} \gamma+\tfrac{1528679}{1536} \pi
   ^2-\tfrac{102232}{315} \log (2)+\tfrac{729}{7} \log (3)\big]\hat{a}+\big[\tfrac{6121}{162}+\tfrac{21}{1024} \pi ^2\big] \hat{a}^3, \nonumber \\ 
c_{6.5}^{\rm ln} &=-\tfrac{1868 }{35}\hat{a}, \nonumber \\ 
c_{7} &= -\tfrac{149628163}{18900}+\tfrac{7628}{21} \gamma +\tfrac{297761947}{393216} \pi
   ^2-\tfrac{1407987}{524288} \pi ^4+\tfrac{4556}{21} \log (2)+\tfrac{12879 }{35}\log (3)+\tfrac{1284 }{25} \pi \hat{a} \nonumber \\
   &\quad+
   \big[-\tfrac{969713}{225}-\tfrac{152 }{3} \gamma+\tfrac{4906229}{12288} \pi ^2-\tfrac{1528
   }{15}\log (2)\big] \hat{a}^2-\tfrac{20 }{3}\hat{a}^4, \nonumber \\ 
c_{7}^{\rm ln} &=\tfrac{3814}{21}-\tfrac{76 }{3}\hat{a}^2, \nonumber \\ 
c_{7.5} &= -\tfrac{113411 }{22050} \pi+
   \big[-\tfrac{3715435931}{28350}-\tfrac{995212}{2835} \gamma +\tfrac{48197747581}{3538944} \pi
   ^2-\tfrac{7009733}{524288} \pi ^4+\tfrac{540788}{2835} \log (2)-\tfrac{4617}{7} \log
   (3)+\tfrac{16}{5} \psi
   ^{\{0,1\}}(\hat{a}) \nn \\
   &\quad-\tfrac{16}{5} \psi ^{\{0,2\}}(\hat{a})\big]\hat{a}+ \big[\tfrac{148627}{450}+\tfrac{136  }{5} \gamma
 -\tfrac{164339}{12288} \pi ^2+\tfrac{392 }{5}\log (2)-24 \log (\kappa)-\tfrac{12}{5} \psi
   ^{\{0,1\}}(\hat{a})-\tfrac{48}{5} \psi ^{\{0,2\}}(\hat{a})\big]\hat{a}^3+\tfrac{2 }{5}\hat{a}^5, \nonumber \\ 
c_{7.5}^{\rm ln} &= -\tfrac{497606 }{2835}\hat{a}+\tfrac{8 }{5}\hat{a}^3, \nonumber \\ 
c_{8} &= \tfrac{403109158099}{9922500}-\tfrac{74909462}{70875} \gamma +\tfrac{3424}{25} \gamma
   ^2+\tfrac{164673979457}{353894400} \pi ^2-\tfrac{160934764317}{335544320} \pi
   ^4+\tfrac{340681718}{1819125} \log (2)+\tfrac{869696
 }{1575}   \gamma \log (2) \nn \\
   &\quad+\tfrac{58208 }{105}\log ^2(2)-\tfrac{199989}{352} \log
   (3)-\tfrac{9765625}{28512} \log (5)-\tfrac{1344}{5} \zeta (3)-\tfrac{3207503  }{33075} \pi \hat{a}\nn\\
&\quad+
   \big[-\tfrac{40220568253}{132300}-\tfrac{4996 }{9} \gamma+\tfrac{2015707491}{65536} \pi
   ^2-\tfrac{724004}{945} \log (2) -\tfrac{4617}{14} \log (3)-\tfrac{32}{5} \log (\kappa
   )-\tfrac{32}{15}
   \psi ^{\{0,1\}}(\hat{a})-\tfrac{16}{15} \psi ^{\{0,2\}}(\hat{a})\big]\hat{a}^2\nn \\
   &\quad+
   \big[-\tfrac{1075453}{24300}-\tfrac{16 }{5} \gamma+\tfrac{679}{1024} \pi ^2-\tfrac{16 }{5}\log
   (2)-\tfrac{16}{5} \log (\kappa ) +\tfrac{8}{5} \psi ^{\{0,1\}}(\hat{a})-\tfrac{16}{5} \psi ^{\{0,2\}}(\hat{a})\big]\hat{a}^4 , \nonumber \\ 
c_{8}^{\rm ln} &= -\tfrac{37454731}{70875}+\tfrac{3424 }{25} \gamma-\tfrac{12634 }{45}\hat{a}^2-\tfrac{16
   }{5}\hat{a}^4+\tfrac{434848}{1575} \log (2), \nonumber \\ 
c_{8}^{\rm ln^{2}} &=\tfrac{856}{25}, \nonumber
\end{align}
\end{widetext}
where $\gamma$ is Euler's constant, $\zeta(n)$ is the Riemann zeta function, $\psi^{\{n,k\}} (\hat{a}) \equiv \psi^{(n)}(\tfrac{i k \hat{a}}{\kappa}) + \psi^{(n)}(\tfrac{-i k \hat{a}}{\kappa}) = 2 \Re[\psi^{(n)}(\tfrac{i k \hat{a}}{\kappa})]$ 
and $\psi^{(n)}(z) = \frac{d^{n+1}}{dz^{n+1}}\ln \Gamma(z)$ is the polygamma function. 

These are partially confirmed by the output of method~II of Table \ref{tab:1}, which provides the same expansion, with at each PN order a Taylor expansion in small~$\hat{a}$.

\subsection{The PN expectation} 

In a two-body system $(m_1,S_1)$ and $(m_2,S_2)$, the precession frequency of the body 1, 
$\Omega_{\rm prec}=\Omega_1$, can be computed following Ref. \cite{Blanchet:2012at} (see Eq. (4.9c) there) 
in terms of the dimensionless binding energy ${\mathcal E}\equiv (E_{\rm system}-M)/\mu $ and angular momentum ${\mathcal L}\equiv L_{\rm system}/(M\mu)$ of the system,  as
\beq
\label{Omega1def}
\Omega_1 = \mu\frac{\partial({\mathcal E}-M\Omega {\mathcal L})}{\partial S_1}\,,
\eeq
where ${\mathcal E}$ and ${\mathcal L}$ are considered as functions of $(m_1,m_2,\Omega, S_1,S_2)$.
Introducing the dimensionless frequency variable $x=(M\Omega)^{2/3}$, Eq. \eqref{Omega1def} implies
\beq
m_2 \Omega_1=m_2  M  \nu \frac{\partial ({\mathcal E}(x)-x^{3/2}{\mathcal L}(x))}{\partial S_1}\,,
\eeq
where the invariant expressions for ${\mathcal E}(x)$ and ${\mathcal L}(x)$ follow straightforwardly from 
Eqs. (5.2) and (5.4) of Ref. \cite{Levi:2015uxa} (concerning the lastly known next-to-next-to-leading order in spin terms; 
for lower order terms see, e.g., Ref. \cite{Blanchet:2012at}).

We list below the resulting expressions for the quantity $\widetilde {E}(x)={\mathcal E}(x)-x^{3/2}{\mathcal L}(x)$, that is
\beq
\widetilde E(x)=\widetilde {E}^{\rm O}(x)+\widetilde {E}^{\rm S}(x)+\widetilde {E}^{\rm SS}(x)+\widetilde {E}^{\rm SSS}(x)+\widetilde {E}^{\rm SSSS}(x)\,,
\eeq
with
\begin{align}
&\widetilde{E}^{\rm O}(x)=\nn\\
&\quad-\frac32  x
+\left(-\frac{9}{8}-\frac{1}{8}\nu\right) x^2
+\left(\frac{19}{16}\nu-\frac{27}{16}-\frac{1}{48}\nu^2\right)x^3
\nn\\
&\quad+\left(\frac{6889}{384}\nu-\frac{405}{128}-\frac{31}{64}\nu^2-\frac{7}{3456}\nu^3-\frac{41}{64}\nu\pi^2\right)x^4\nonumber\\
&
\quad+\bigg(-\frac{1701}{256}+\frac{451}{384}\nu^2\pi^2+\frac{43}{1152}\nu^3-\frac{24689}{3840}\nu \nn \\
&\quad
-\frac{71207}{2304}\nu^2+\frac{11}{20736}\nu^4+\frac{64}{5}\nu\gamma+\frac{128}{5}\nu\ln(2)\nonumber\\
&\quad
+\frac{32}{5}\nu\ln(x)+\frac{1291}{1024}\nu\pi^2\bigg)x^5+\mc{O}(x^6)
\,,\nonumber\\
&\widetilde{E}^{\rm S}(x)= 
\left[\left(-\Delta-\frac{1}{2}\nu+1\right)\chi_1
+\left(1-\frac{1}{2}\nu+\Delta\right)\chi_2\right]x^{5/2}\nonumber\\
&
\quad+\bigg[\left(\frac{3}{2}+\frac{31}{48}\nu\Delta-\frac{3}{2}\Delta+\frac{1}{24}\nu^2-\frac{121}{48}\nu\right)\chi_1 \nn \\
&\quad
+\left(\frac{3}{2}\Delta+\frac{3}{2}+\frac{1}{24}\nu^2-\frac{121}{48}\nu-\frac{31}{48}\nu\Delta\right)\chi_2\bigg]x^{7/2}\nonumber\\
&\quad+\left[\bigg(-\frac{373}{32}\nu-\frac{27}{8}\Delta+\frac{211}{32}\nu\Delta+\frac{43}{12}\nu^2-\frac{7}{48}\nu^2\Delta+\frac{1}{48}\nu^3 \right.\nn \\
&\quad+\frac{27}{8}\bigg)\chi_1
+\bigg(\frac{1}{48}\nu^3-\frac{373}{32}\nu+\frac{27}{8}+\frac{27}{8}\Delta+\frac{7}{48}\nu^2\Delta \nn \\
&\quad\left.-\frac{211}{32}\nu\Delta+\frac{43}{12}\nu^2\bigg)\chi_2\right]x^{9/2}+\mc{O}(x^{11/2})
\,,\nonumber\\
&\widetilde{E}^{\rm SS}(x)= \bigg[  \left(\frac14 \Delta-\frac14+\frac12 \nu\right)\chi_1^2+\bigg(-\frac14 \Delta-\frac14+\frac12 \nu\bigg)\chi_2^2 \nn \\
&\quad-\chi_1\chi_2\nu \bigg] x^3
+\bigg[  \left(\frac{13}{24}\Delta-\frac{7}{12}\nu^2-\frac{29}{24}\nu\Delta-\frac{13}{24}+\frac{55}{24}\nu\right) \nn \\
&\quad\times\chi_1^2+\left(-\frac16 \nu^2-\frac12 \nu\right)\chi
_2\chi_1\nonumber\\
&\quad\left.
+\left(-\frac{13}{24}\Delta+\frac{55}{24}\nu+\frac{29}{24}\nu\Delta-\frac{7}{12}\nu^2-\frac{13}{24}\right)\chi_2^2\right]x^4\nonumber\\
&
\quad+\bigg[\bigg(\frac{3095}{288}\nu+\frac{607}{288}\nu^2\Delta+\frac{59}{144}\nu^3+\frac{67}{32}\Delta -\frac{3017}{288}\nu^2\nn \\
&\quad-\frac{67}{32}-\frac{1889}{288}\nu\Delta\bigg)\chi_1^2+\left(\frac{53}{72}\nu^2-\frac{15}{8}+\frac{143}{24}\nu\right)\nu\chi_2\chi_1\nonumber\\
&
\quad+\bigg(\frac{3095}{288}\nu-\frac{3017}{288}\nu^2-\frac{607}{288}\nu^2\Delta+\frac{59}{144}\nu^3-\frac{67}{32}-\frac{67}{32}\Delta \nn \\
&\quad+\frac{1889}{288}\nu\Delta\bigg)\chi_2^2\bigg]x^5 +\mc{O}(x^6)
\,,\nonumber\\
&\widetilde{E}^{\rm SSS}(x)= 
\left[\left(\frac{1}{2}\nu-\frac{1}{4}+\frac{1}{4}\Delta\right)\chi_1^3
+\left(-\frac{1}{4}+\frac{1}{4}\Delta-\frac{1}{2}\nu\right)\chi_2\chi_1^2\right.\nonumber\\
&\left.
+\left(-\frac{1}{4}-\frac{1}{2}\nu-\frac{1}{4}\Delta\right)\chi_2^2\chi_1
+\left(-\frac{1}{4}-\frac{1}{4}\Delta+\frac{1}{2}\nu\right)\chi_2^3\right]\nu x^{9/2}+\mc{O}(x^{11/2})
\,,\nonumber\\
&\widetilde{E}^{\rm SSSS}(x)= \mc{O}(x^6)  \,.
\end{align}

In the previous expressions we have replaced the spin variables $S_{1,2}$ by their dimensionless counterparts $\chi_{1,2}=S_{1,2}/m_{1,2}^2$.
In order to compare with the GSF expression derived above, we compute the spin precession invariant $\psi=(m_2 \Omega_1)/y^{3/2}$, where the variable $y$ is related to $x$ by $x=(1+q)^{2/3}y$.
Linearizing in $q$ we find 
\begin{eqnarray}
\psi(y)&=& \frac{3}{2}y+\frac{9}{8}y^2+\frac{27}{16}y^3\nn\\
&+&\left(-y^{3/2}-\frac12  y^{5/2}-\frac{15}{8} y^{7/2}\right)\chi_2-\frac12 y^3\chi_2^2\nn\\
&+&\nu \delta\psi(y) + \mc{O}(\nu^2,y^4)\,,
\end{eqnarray}
with
\begin{align}
\delta\psi(y) &= y^2-3 y^3 +\left(y^{3/2}+\frac{16}{3}y^{7/2}\right)\chi_2\\
&\quad+ \left(-y^{3/2}+\frac32 y^{5/2}+\frac{9}{8} y^{7/2}-2y^3\chi_2\right)\chi_1+ \mc{O}(y^4)\,.\nn
\end{align}
Here the zeroth-order in $\nu$ contribution to $\delta \psi$ coincides with the Kerr value (see, e.g., Eq. (70) of Ref. \cite{Iyer:1993qa}); the $\mathcal{O}(\nu)$ Schwarzschild contribution to $\delta \psi$ coincides with previous results \cite{Bini:2014ica,Dolan:2013roa}; the first terms linear in spin in $\delta \psi$ agree with our first-order GSF result \eqref{Eq:deltapsi8pn}.

\subsection{Numerical results}

\begin{figure}[t]
	\includegraphics[width=\columnwidth]{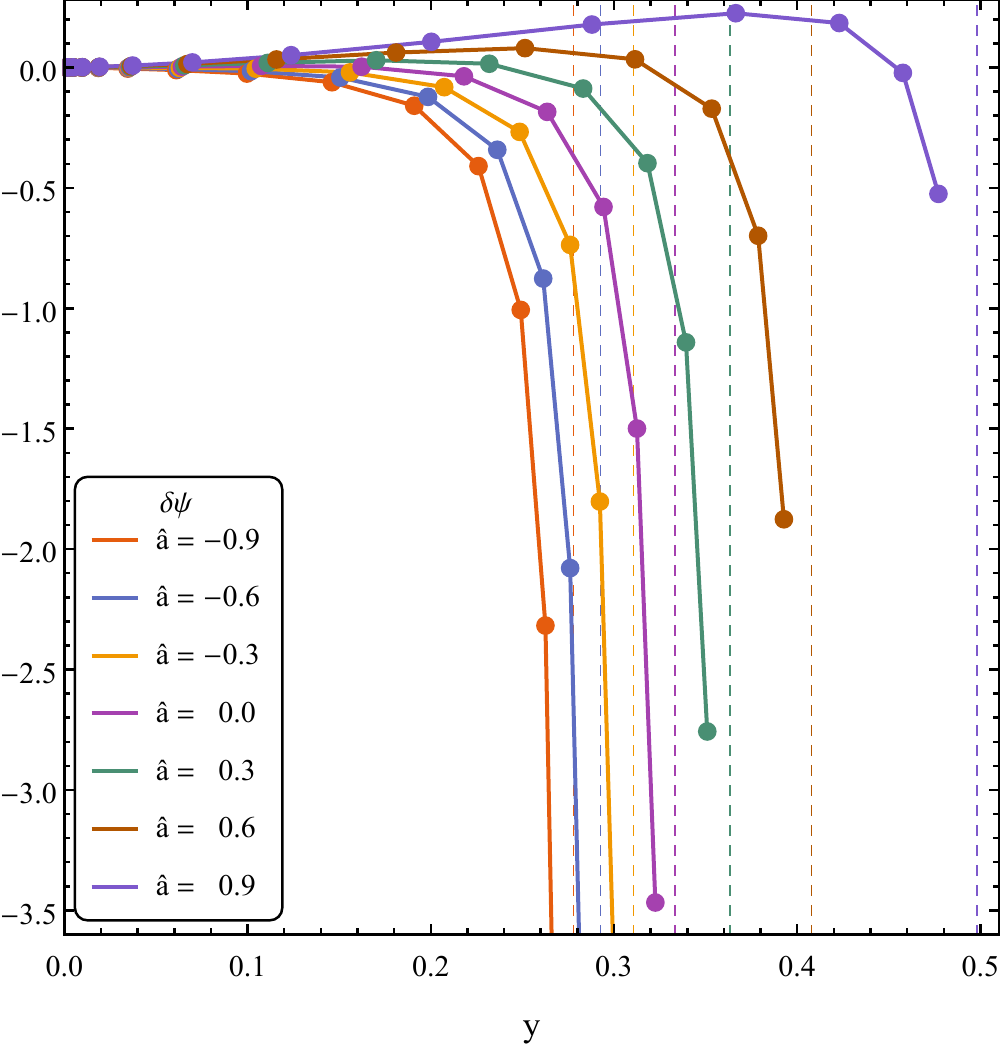}
	\caption{Numerical results for $\delta\psi$ for various values of the spin $a$. The vertical dashed lines show the location of the lightring for that value of the spin, where $\delta\psi$ diverges.}
	\label{fig:numdataplot}
\end{figure}

\begin{figure}[t]
    \includegraphics[width=\columnwidth]{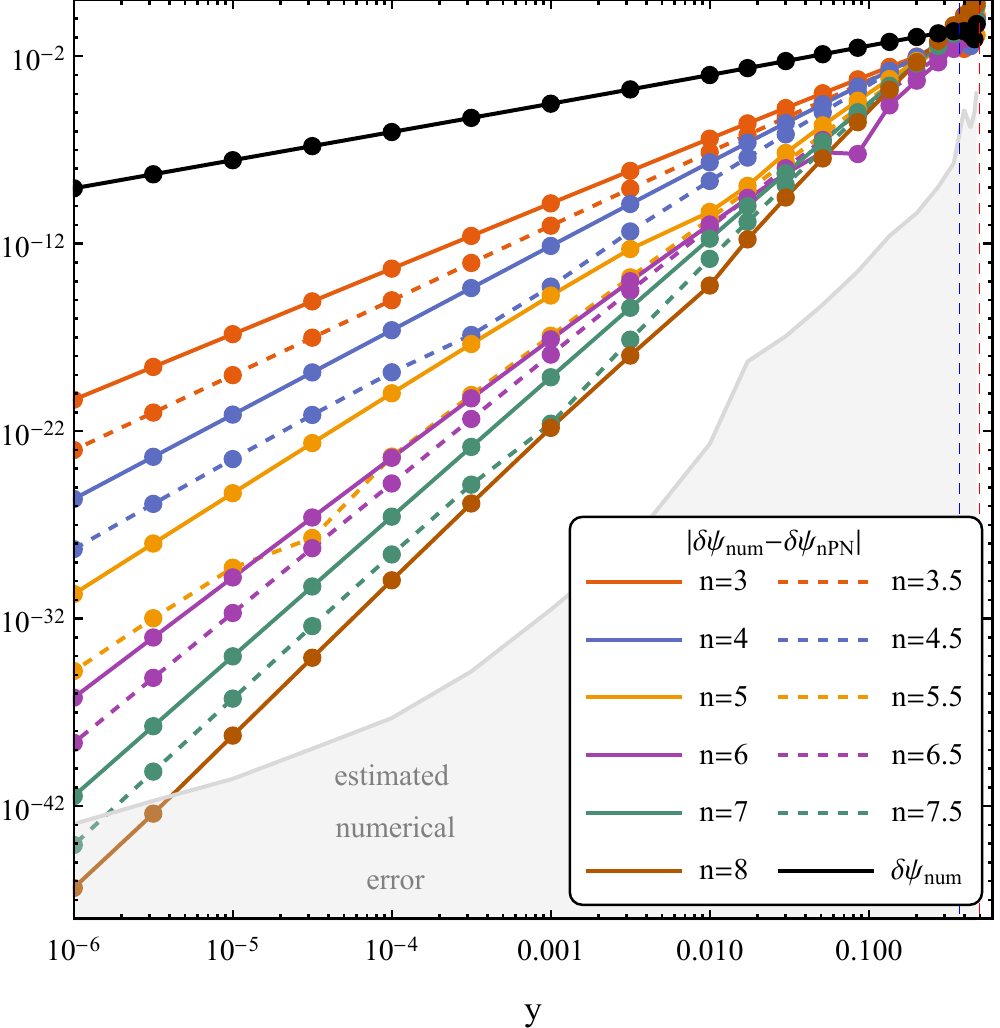}
	\caption{Comparison of the numerical and PN results for $\delta\psi$ in the weak field regime for $\hat{a} = 0.9$. The plotted lines show the residual after subtracting the $n$th order PN approximation. The slopes of each line is compatible with an $(n+1/2)th$ order residual, as one would expect. The blue and red vertical dashed lines show the location of the ISCO and the lightring respectively. The shade region gives an estimate on the numerical error in the calculation of $\delta\psi_{\mathrm{num}}$}
	\label{fig:weakfieldcomp}
\end{figure}

\begin{figure}[t!]
	\includegraphics[width=\columnwidth]{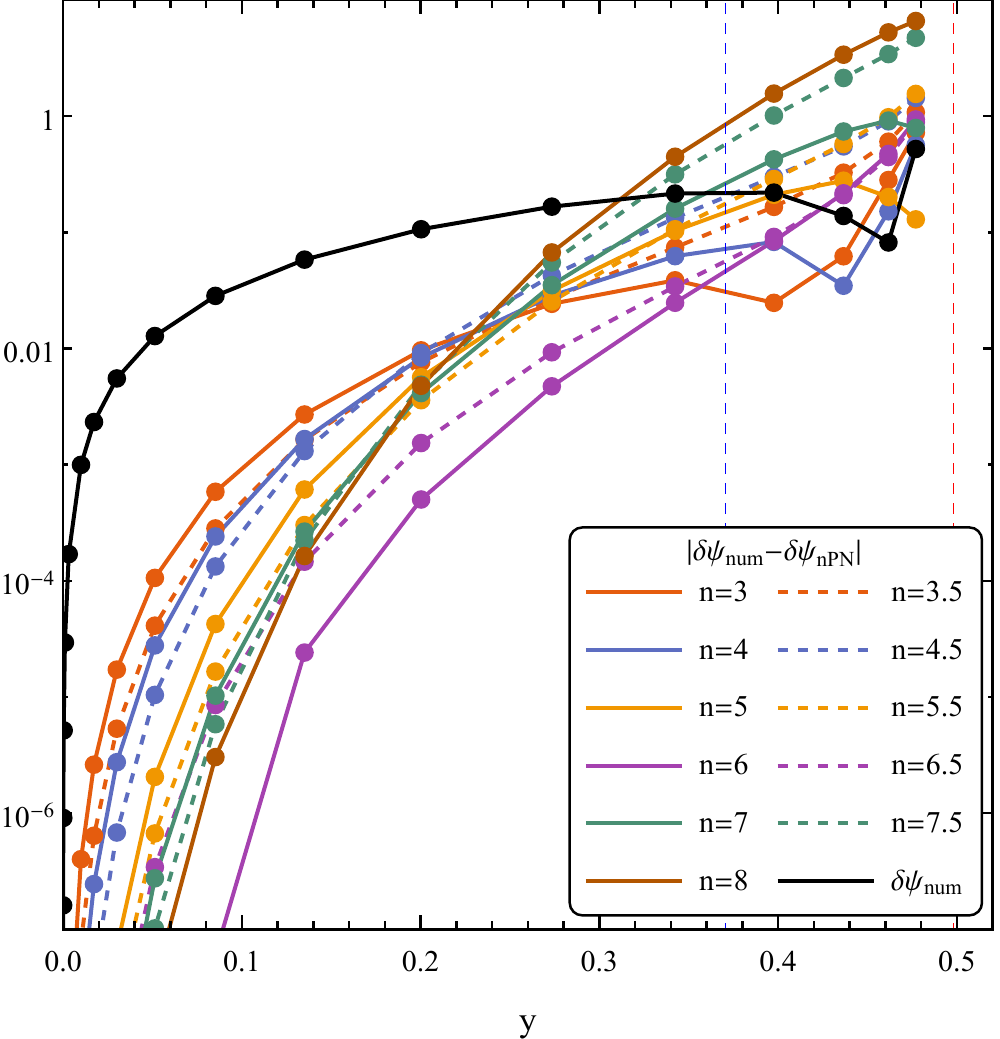}
	\caption{Comparison of the numerical and PN results for $\delta\psi$ in the strong field regime for $\hat{a} = 0.9$. Above $y\approx 0.23$ the PN residuals no longer consistently improve with higher PN order, demonstrating the asymptotic nature of the PN expansion. The relatively good performance of the $6PN$ approximant seems mostly coincidental due to a zero crossing. The blue and red vertical dashed lines show the location of the ISCO and the lightring respectively.}
	\label{fig:strongfieldcomp}
\end{figure}

Method III obtains high accuracy numerical results for $\delta\psi$ without any post-Newtonian assumptions. Fig.~\ref{fig:numdataplot} shows the results for a variety of spins. One obvious feature is that as the (unstable) circular orbits approach the lightring $\delta\psi$ diverges. This behaviour is well-known in the analogous case of the redshift invariant \cite{Barack:2011ed,Akcay:2012ea,Bini:2015xua}, and was studied in the case of  $\delta \psi$ around Schwarzschild  in \cite{Bini:2014zxa},  which concluded that the light-ring divergence of $\delta \psi$  is proportional to $E^2$, where $E$ is the orbital energy. The data here is also compatible with a divergence~$\propto E^2$. The full numerical results are available from the \emph{black hole perturbation toolkit} website \cite{BHPToolkit}.

Fig.~\ref{fig:weakfieldcomp} shows a comparison of the numerical results with the obtained PN results in the weak field regime. Shown are the residuals after subtracting successive orders in the PN expansion. We see a consistent improvement in the weak field, providing a strong verification of both the analytical PN results and the numerical results.

Fig.~\ref{fig:strongfieldcomp} shows the same plot but with a focus on the strong field regime. Here the picture is very different. Around $y\approx 0.24$ we observe a locus where all PN approximants do about equally well (with the 6 and 6.5 PN terms as notable exceptions). Above this there is no noticeable improvement from going to higher PN orders.

\section{Discussion and outlook}

In this paper we have, for the first time, calculated the GSF corrections to the spin precession invariant along circular equatorial geodesic orbits in a perturbed Kerr spacetime, generalizing previous results limited to the case of a perturbed Schwarzschild spacetime. This calculation has been done with a variety of methods and techniques providing ample cross-validation.

Comparison with existing PN results using the first law of binary mechanics \cite{Blanchet:2012at}, provides a strong validation of the used radiation gauge GSF techniques employed here, while also validating the previous PN results.

Cross validation between the different GSF calculations, which vary in the level of rigor in their derivation, validates some of the underlying assumptions. In particular, a subtle importance is the agreement we find between the methods despite the differences in harmonic projections. State of the art radiation gauge self-force codes project from spin-weighted spheroidal harmonics to scalar spherical harmonics to meet up with rigorously defined regularization techniques, which has a large negative impact on the computational costs. In this work we have shown agreement between such a projected numerical code, and an unprojected analytical code without needing and additional correction terms. Investigating if such agreements between projections hold in more generic orbital configurations or for gauge dependent quantities (such as the self-force itself) would be of great importance in developing more efficient numerical codes for realistic self-force models.

An important application of the results in this paper will be to inform effective-one-body (EOB) theory \cite{Buonanno:1998gg,Damour:2001tu}. As shown in \cite{Bini:2014zxa}, the spin precession can be 
used to determine contributions to the eﬀective-one-body Hamiltonian for spinning black holes 
relating to the secondary spin. This transcription will be left to future work.

This work, focusing on circular equatorial orbits, is a first step in determining the spin precession around Kerr black holes. The formalism for extending this work to eccentric equatorial orbits has already been laid out \cite{Akcay:2017azq} and should provide a basis for generalizing to generically inclined orbits. This should provide additional avenues of cross-validating difficult GSF calculations and informing EOB.

\section*{Acknowledgments}

DB thanks ICRANet and the italian INFN for partial support and IHES for warm hospitality at various stages during the development of the present project. MvdM was supported by European Union's Horizon 2020 research and innovation programme under grant agreement No 705229. The numerical results in this paper were obtained using the IRIDIS High Performance Computing Facility at the University of Southampton.

\bibliography{gyro_prec_kerr,commongsf,meent,journalshortnames}

\end{document}